\pdfoutput=1 
\input{aipcheck}
\documentclass[
    ,final            % use final for the camera ready runs
    ,numberedheadings % uncomment this option for numbered sections
  ]
  {aipproc}

\layoutstyle{8x11double}

\begin{document}

\thispagestyle{empty}

%\title{Search for the MSSM Higgs bosons at the LHC}
%\title{Bump Hunting for Final States with two Invisible particles at proton-proton colliders}

\title{Searching in 2-Dimensional Mass Space for Final States with 2 Invisible Particles}


\author{G.~Anagnostou }{
  address={Institute of Nuclear and Particle Physics, NCSR Demokritos, Greece }
}

\begin{abstract}
\\

A method to search for particles of unknown masses  in final states with two invisible particles is presented.
Searching for final states with missing energy is a challenging task usually performed in the tail of a missing energy related distribution. 
%The similarity of signal and background shapes, makes the discovery a difficult task and gives no hint for the nature of new physics except that it exists.
The search method proposed is based on a 2-Dimensional mass reconstruction of the final state with two invisible particles. 
Thus, a bump hunting is possible, allowing a stronger signal versus background discrimination.
%The method is applicable in many topologies  predicted for dark matter candidates.
Parameters of the BSM theory can be extracted from the mass distributions, a valuable step towards understanding its true nature.
%The method is based on searching in  the 2-Dimensional mass space of two unknown particles for the point for which the event is more likely to originate from a proton-proton collision. The probability is calculable for all solvable points in the 2D mass plane as in this case a kinematic reconstruction is possible. Using the PDFs we can then assign a different probability to each point. 
The proof of principle of the method is based on the existing SM top pairs in their dilepton final state. 
Many interesting topologies including dark matter candidates or heavy top partners can be searched this way at the LHC.
\end{abstract}

\maketitle

%%%%%%%%%%%%%%%%%%%%%%%%%%%%%%%%%%%%%%%%%%%%
%% MAINMATTER
%%%%%%%%%%%%%%%%%%%%%%%%%%%%%%%%%%%%%%%%%%%%

%======================================================================================

\section{Introduction}

After Higgs boson discovery, it is not easy to figure out which experimental signatures are the most promising.
%and can show to the next step in particle physics. 
%Missing energy final states are an interesting case, predicted by well motivated BSM theories ~\cite{susy,extradimensions,littlehiggs1}.
Missing energy final states are an interesting case, predicted by well motivated BSM theories ~\cite{susy}-\cite{littlehiggs1}.
%An interesting case are missing energy final states predicted by well motivated BSM theories.
%Final states with large missing energy are predicted by well motivated BSM theories ~\cite{bsm1-bsm3}.
The energy missing originates from invisible particles such as neutrinos (e.g heavy top partners) or WIMPS as dark matter candidates.

%Many well motivated BSM theories predict final states with large missing energy originating from invisible particles such as neutrinos (e.g heavy top partners) or WIMPS as dark matter candidates. 

In the case of Higgs boson, the Standard Model (SM)  was predicting everything except its mass: the cross-section, its decay channels, the couplings to other particles etc.
Even the Higgs mass was  predicted indirectly by the electroweak fit,
so that the search could be narrowed to a few tens of GeV of the invariant mass spectrum ~\cite{electroweakfit}.
The LHC machine and detectors were actually designed for Higgs boson discovery.
Based on accurate Monte-Carlo predictions the search could be tuned 
to extract the signal easier. For example, multivariate methods were trained to
discriminate signal from background processes based on simulated Higgs events. 
Without this tuning in the design of the experiment/detectors/analysis, Higgs discovery
would probably be harder to be established and have taken longer time.
In addition, the discovery was based on invariant mass observables,  in channels
with visible decay products (photons, electron/muons). It was a bump hunt, a case where the
shape of signal and background processes is different, so that a robust discovery is
easier.

\begin{figure}[t!]
%\centering
\includegraphics[scale=0.091]{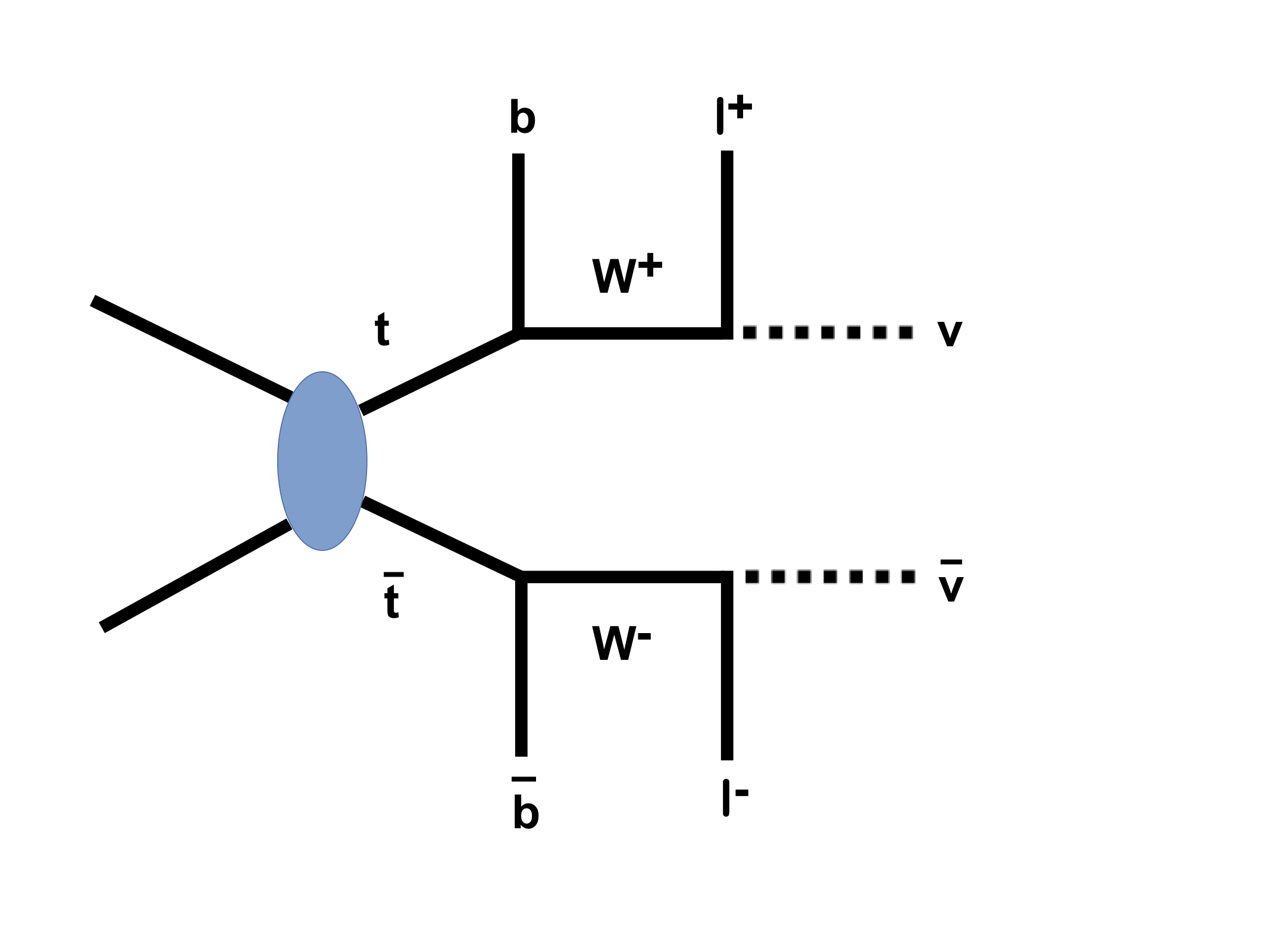}
\includegraphics[scale=0.091]{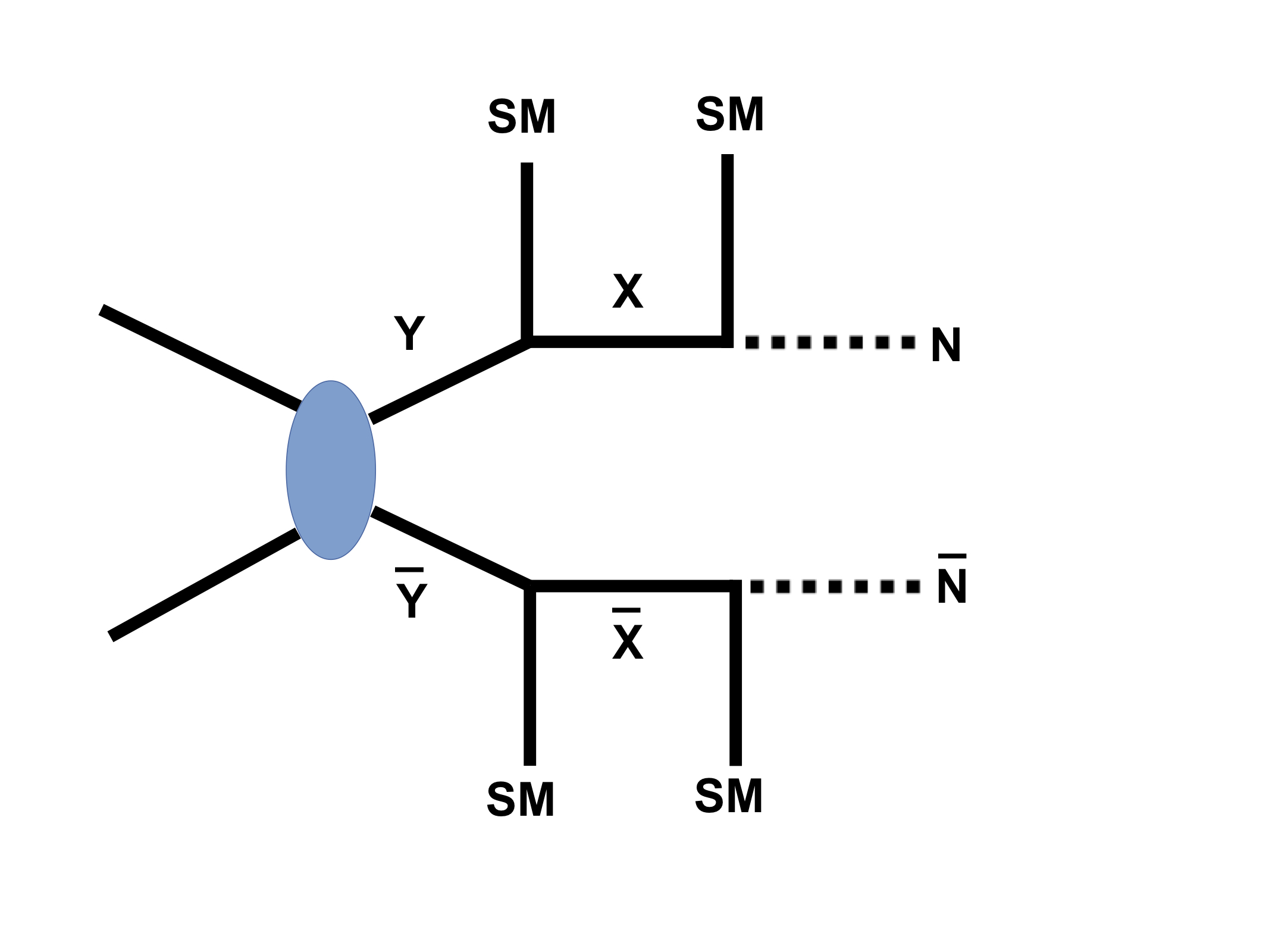}
\caption{The existing SM top pair decay with two invisible particles (left) and  a similar pair produced generic BSM topology (right).}
\label{fig:feynman}
\end{figure}

Searching for new physics in final states with missing energy is a challenging task and
certainly much harder than searching for a particle with known properties such as the SM Higgs boson ~\cite{metsearches1} \cite{metsearches2}.  
%A symmetric topology with missing energy originating from two invisible particles can be seen in \mbox{Figure \ref{fig:feynman}} (left), and the existing similar SM dilepton top pairs in Figure \ref{fig:feynman} (right).
An existing SM diagram with two invisible particles can be seen in  Figure \ref{fig:feynman} together with a similar pair produced generic  BSM topology.
Understanding the physics is much easier in possible BSM signals with visible
particles. In this case, the invariant  masses can be reconstructed  up to  combinatoric ambiguities.
In final states with invisible particles, instead of a bump hunt in a small region of the invariant mass spectrum, the search has to be performed using the tail of a  missing energy related distribution.
Due to the small difference in shape between signal and background distributions the establishment of a discovery in this case is a difficult problem. 
%In this case, the establishment of a discovery is a difficult problem due to the small difference in shape between signal and background distributions.
In addition, any mismeasurement or mismodelling of any type of the physics objects used can introduce a fake missing energy signal.
Even when a discovery is established, no additional information about the  new physics would be available except that it exists.

The classical way to search for new particles is by applying a set of selection rules (cuts) which
are usually targeting the region of phase space where signal dominates. This depends strongly on
the specific model and usually selects rectangular phase space regions of the energy and transverse
momenta of the reconstructed objects. Sometimes these kinematic observables are used as input to
multivariate discriminators like likelihoods, boosted decision trees and neural networks. It is
common that experiments tune their cuts/multivariate observables such as to be optimized/trained for a
specific choice of the BSM model parameters. If new physics is not in the parameter space region for which
the search was optimized/trained, it is likely that the sensitivity of the search will be reduced.
So it is highly desirable that the performance of the search does not depend
on the model or that it is as model independent as possible.

The mass and spin are the most important characteristics of elementary particles according to QFT.
The mass space is ideal for resonance searches: events are concentrated in a small region, whereas background events have no reason to do the same.
Mass observables do not require optimization or training. 
Ideally, searches should be performed in multidimensional mass spaces with as many dimensions as the number of unknown particles.
They are commonly used in searches with visible particles and are also used in topologies with a single invisible particle.
This paper proposes to use them extensively in final states with two invisible particles, as there are  many interesting applicable topologies and can offer all the advantages of mass observables.

In the next sections, the method to perform 2-Dimensional mass reconstruction in final states with two invisible particles is described. 
Initially, (section 2) the generation and simulation of the signal and background processes are discussed.
The next section concerns the description of the method.
Section 4 has several topologies and applications starting with the dilepton top pairs as a proof of principle.
%without any a-priory knowledge of the masses, top quark and W boson can be observed as a peak in the 2-Dimensional mass plane.
Then, in the next example, a generic topology of anything decaying like dilepton top pairs is presented ($\mathrm{pp \rightarrow T'\bar{T'} \rightarrow W'b W'b}$).
%In this case, the Y particle is a heavy top partner T' and X is a new heavy gauge boson ($\mathrm{pp \rightarrow T'\bar{T'} \rightarrow W'b W'b }$, with W' decaying as ($\mathrm{W' \rightarrow e, \mu }$ ).
%The search is performed in the  mass plane of the unknown particles T' and W'.
More applications are discussed such  as the search for a pair produced heavy top partner $\mathrm{T'}$ decaying as SM top quark as well as a new heavy neutral gauge boson $\mathrm{Z'}$ decaying to top pairs. 
Finally, the usage of the 2-Dimensional mass reconstruction is proposed for identification of dilepton top pairs in final states with large missing energy as well as for a top mass measurement in the same channel.
The description that follows is based on simulated events with all the complexity available in a fast simulation package. 
%It is worth mentioning that the method has already been applied in CMS Run1 dataset for a generic search for anything decaying like dilepton top pairs \cite{mymethod,cms-b2g-12-025,sarah,beyondtwogenerations}.
It is worth mentioning that the method has already been applied in CMS Run1 dataset for a generic search for anything decaying like dilepton top pairs \cite{mymethod}-\cite{beyondtwogenerations}.

%our knowledge for the parameters of the BSM physics would be significantly enhanced by the measurement of the masses.
%Additionally, in mass space, the discovery can be established in a robust way: the
%reconstructed masses allow a data-driven estimate of the background from the side-bands.
%The existence of the invisible particles in the decay chain makes the extraction of the unpredicted by the
%model masses a difficult problem [4][5]. But the masses are important parameters of the model and
%a valuable input towards the full understanding of the theory. So, if there is a way to reconstruct
%unknown masses in searches with invisible particles, we should do it anyway.

%In the following section, a method that allows  
%It is often believed that reconstruction of  particles masses in
%topologies with two invisible particles is not feasible. In the rest
%of this study a counter example is presented with the simultaneous
%reconstruction of W boson and top quark in the dilepton top-pair final
%state in a p-p collider like LHC. The method can also be used to search
%in a model independent way for a new generation $\mathrm {t'}$ and any resonance decaying
%to top pairs $\mathrm{pp\rightarrow X'\rightarrow t\bar{t}}$.
%In the next section  event simulation and selection is described.
%In section 3 the method is presented together with possible applications.

%%%%%%%%%%%%%%%%%% F I G U R E %%%%%%%%%%%%%%%%%%%%%%%%%%%%%%%%%%%%%%%%%%%%%%
\begin{figure}[t!]
%\centering
%\includegraphics[scale=0.36]{eps/single_event_top_left.eps}
%\includegraphics[scale=0.36]{eps/single_event_top_right.eps}
\includegraphics[scale=0.36]{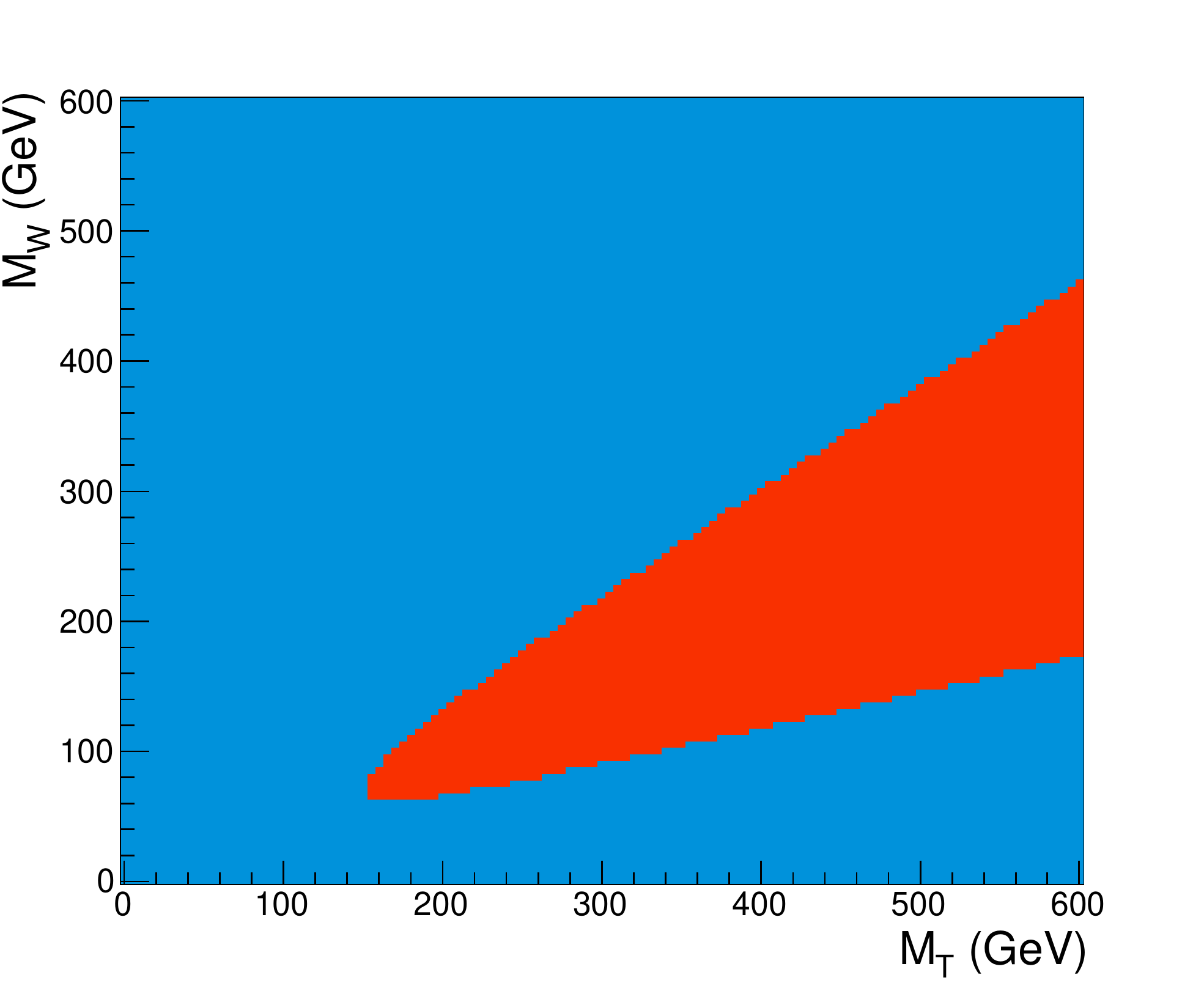}
\includegraphics[scale=0.36]{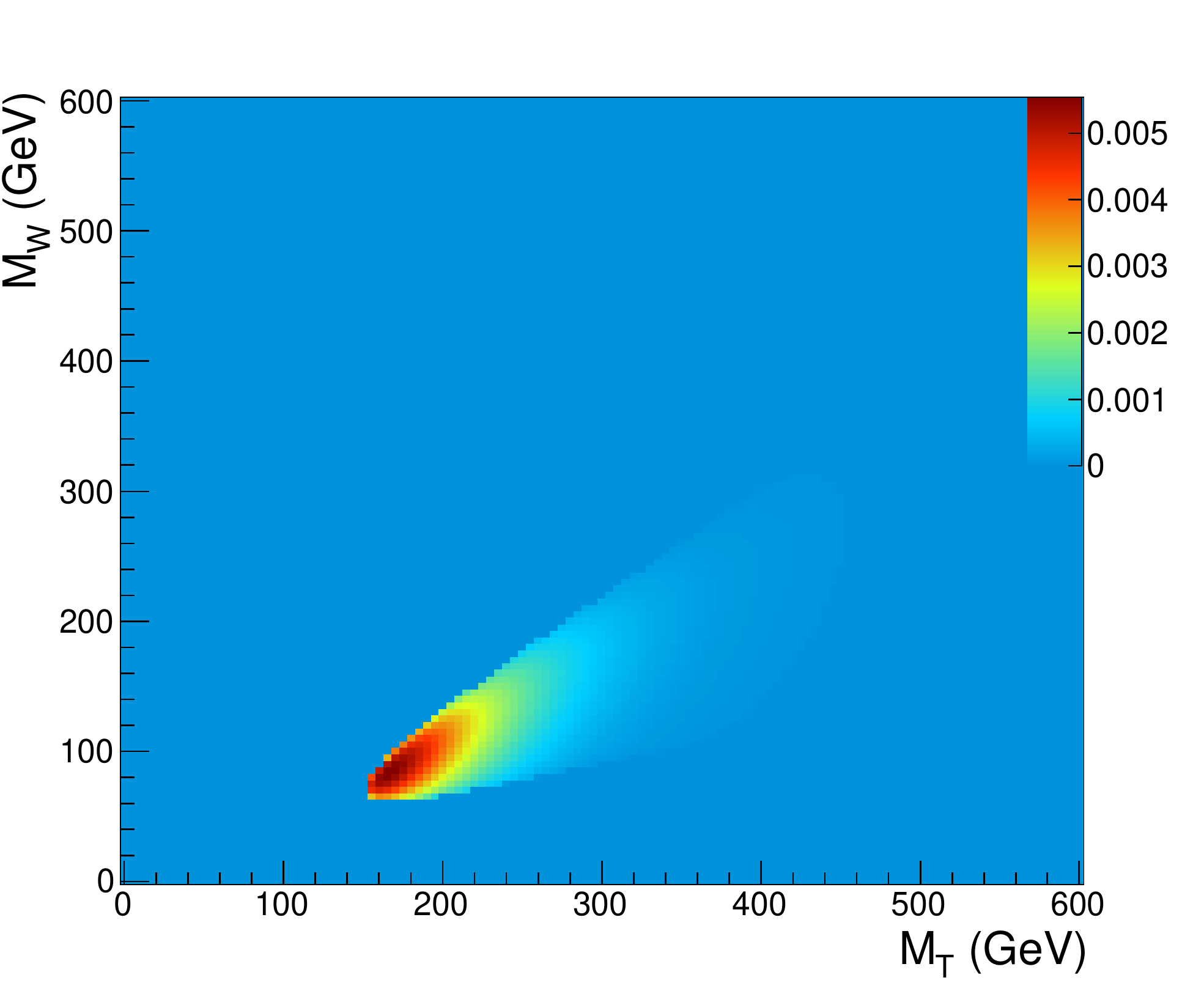}
\caption{(Left): Solvability for one of the possible solutions of  a single top-pair event in the 2-Dimensional mass plane. (Right): Solvability weighted with the PDFs and normalized to unit volume for the same solution and event.}
\label{fig:singleevent_sol}
\end{figure}
%%%%%%%%%%%%%%%%%% F I G U R E %%%%%%%%%%%%%%%%%%%%%%%%%%%%%%%%%%%%%%%%%%%%%%

\section{Event Simulation and Selection}

All signal and background processes were generated and simulated for an intergrated luminosity of 50 $\mathrm{fb^{-1}}$ at an LHC collision energy of 13 TeV.
The SM processes  were  generated using Pythia8 \cite{pythia8}:
top pairs, single boson (Drell-Yan, W+jets),  dibosons ($\mathrm{WW}$, $\mathrm{WZ}$, $\mathrm{ZZ}$) as well as single top events.

Signal events for the generic search of anything decaying like top pairs  ($\mathrm{pp \rightarrow T'\bar{T'} \rightarrow W'b W'b }$) were generated using the littlest Higgs Model  \cite{littlehiggs2} \cite{littlehiggs3}.
An implementation of the latter can be found in  Whizard 2.2.0 event generator \cite{whizard1} \cite{whizard2}. 
The model has both a pair produced heavy top partner  $\mathrm{T'}$ as well as a  new heavy gauge boson $\mathrm{W'}$.
Initially, the hard scattering  ($\mathrm{pp \rightarrow T'\bar{T'}}$) was generated with Whizard.
The decays of both heavy top partner  ($\mathrm{T' \rightarrow W' b}$) and new heavy gauge boson to leptons ($\mathrm{W' \rightarrow l, \nu, l=e, \mu }$)  were performed with Pythia8  using a flat matrix element  in order to create a simplified model.
%A grid  of samples is produced with masses $\mathrm{M_{T'} }$ and $\mathrm{M_{W'} }$ up to 2 TeV in steps of 200 GeV, for points with   $\mathrm{M_{T'}>M_{W'} }$.
Additional signal samples for the application of the method in cases with  a single unknown particle have been produced.
More specifically, for the  pair production of a heavy top partner decaying to SM particles ($\mathrm{pp \rightarrow T'\bar{T'} \rightarrow Wb Wb}$)  signal events were generated with Pythia8.
The same generator was used for the production of a heavy new $\mathrm{Z'}$ decaying to SM top pairs.
%($\mathrm{pp \rightarrow Z' \rightarrow t\bar{t} \rightarrow Wb Wb}$).
All signal and background processes were further processed with Delphes-3.4.0  detector simulation package \cite{delphes} using  the parameters of a typical LHC detector (CMS).

Events were selected by requiring at least two energetic leptons, two energetic jets and missing transverse energy.
The two highest in $\mathrm{P_{T}}$  leptons and jets  were chosen.
Additionally, the two leptons were opposite charged and  identified as electrons or muons with  $\mathrm{P_{T}}>30$ GeV.
The jets were reconstructed using the AK4 algorithm and required to have  $\mathrm{P_{T}}>$ 30 GeV with at least one of them tagged as originating from a bquark.
Jets and leptons were selected in the pseudorapidity region $\mathrm{|\eta|} <$ 2.4 and $\mathrm{|\eta|} <$ 2.1 respectively.
%For the study with top pairs as a signal, events with transverse missing energy less than 100 GeV are rejected.
%For the generic search for anything decaying like dilepton top pairs, the selection is the same for the leptons and jets, with the missing energy requirement raised to 200 GeV.
 Finally, events with transverse missing energy less than 100 GeV were rejected.
Events satisfying the above requirements were used as input to the algorithm described in the next section.

%%%%%%%%%%%%%%%%%% F I G U R E %%%%%%%%%%%%%%%%%%%%%%%%%%%%%%%%%%%%%%%%%%%%%%
\begin{figure}[b!]
\centering
\includegraphics[scale=0.36]{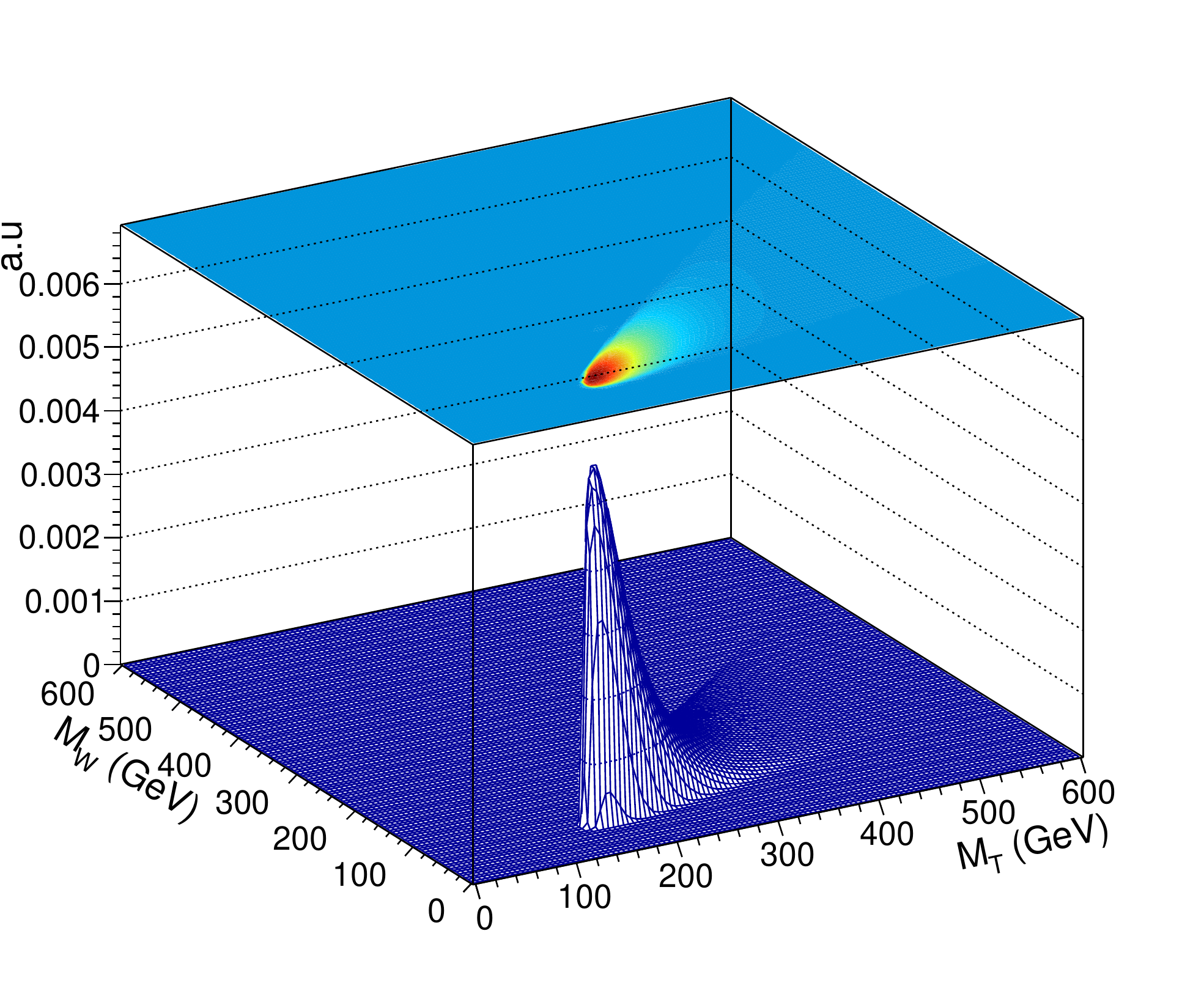}
\caption{Average  PDF weighted solvability for the 100  smeared events produced from the same initial top pair event.
The distribution is  normalized to unit volume.}
\label{fig:singleevent_3D_300}
\end{figure}
%%%%%%%%%%%%%%%%%% F I G U R E %%%%%%%%%%%%%%%%%%%%%%%%%%%%%%%%%%%%%%%%%%%%%%

\section{The Method}

The dilepton top pairs system of equations has an analytical solution \cite{lans1} \cite{lans2}.
The algorithm takes as an input the masses  of top quark and W boson as well as the visible particle's momenta and gives as an output the momenta of the two neutrinos.
For any $\mathrm{M_{T}}$ and $\mathrm{M_{W}}$ values for which the system is solvable, all event kinematics are calculable, including the fractions of the proton's energies participating in the hard process. 
Using these fractions as an input to the Parton Distribution Functions (PDFs), a probability can then be assigned for each point of the 2-Dimensional mass plane.
The one that is more likely to originate from a p-p collision can be chosen as the $\mathrm{M_{T}}$, $\mathrm{M_{W}}$ estimation for the event.

More specifically, the kinematics of $\mathrm{ t\bar{t}}$ dilepton events can be expressed by two linear and six non linear equations (Appendix).
The system is solvable with respect to the unknown neutrino and antineutrino momenta, provided that the masses $\mathrm{M_{T}}$, $\mathrm{M_{W}}$, the  momentum of bquarks and leptons as well as the missing energy components are available.
Each possible input can give 0, 2 or four different solutions for the  unknown neutrino and antineutrino momentum components.
In addition, there are two possible combinations of bjets and leptons that could originate from the same top quark, giving
in total up to eight solutions for  specific $\mathrm{M_{T}}$ and  $\mathrm{M_{W}}$ values. 
Knowledge of the momenta of the invisible neutrinos allows full kinematic reconstruction of the event including the four-vectors of W bosons, top quarks and the $\mathrm{t\bar{t}}$ system.

%%%%%%%%%%%%%%%%%% F I G U R E %%%%%%%%%%%%%%%%%%%%%%%%%%%%%%%%%%%%%%%%%%%%%%
\begin{figure}[t!]
\centering
\includegraphics[scale=0.41]{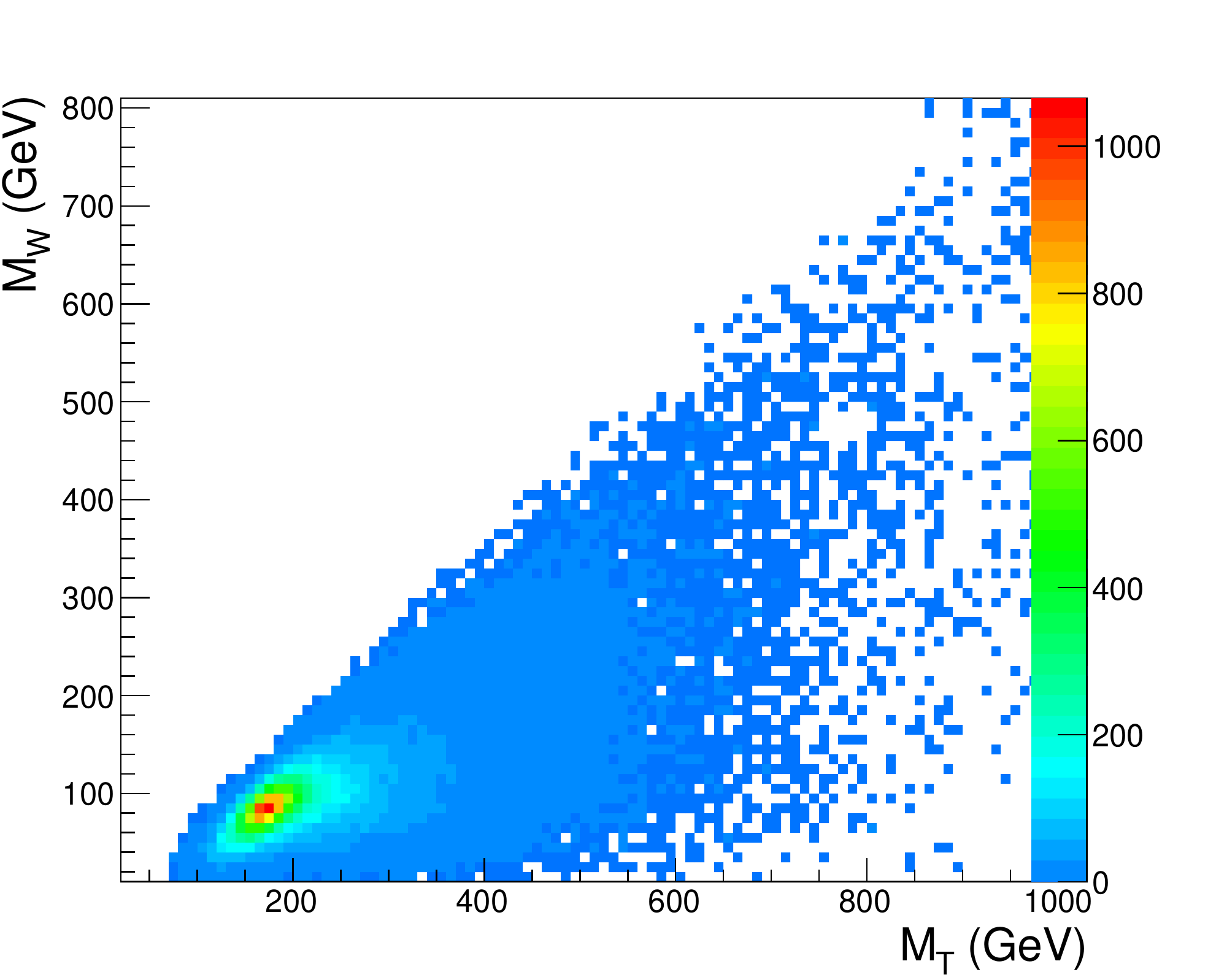}
\caption{Reconstructed $\mathrm{M_{T}}$ and  $\mathrm{M_{W}}$  per event for a sample of signal (top-pairs)  and background events corresponding to an intergrated luminosity of 50 $\mathrm{fb^{-1}}$ at  13 TeV.}
\label{fig:2D_toppairs}
\end{figure}
%%%%%%%%%%%%%%%%%% F I G U R E %%%%%%%%%%%%%%%%%%%%%%%%%%%%%%%%%%%%%%%%%%%%%%

Searching for topologies with two invisible particles requires no a-priori knowledge of the masses, as this should be the result rather than the input.
It is the inverse problem with respect to the analytical solution: given the visible particles and the topology we are looking for the unknown masses per event.
In order to solve the inverse problem, every point of the   $\mathrm{M_{T}}$, $\mathrm{M_{W}}$  plane is tested for possible solutions.
%An event is considered solvable if any of the two possible combinations of bjets and leptons is solvable.
The  mass plane  can be scanned  in steps of a few GeV (in this case 5 GeV) to produce the area in which each one
of the eight possible solutions exists or not.
The existance of a real solution makes the event solvable for this specific mass point.
%Solvability for a single event can be defined as  the existance (or not) of a specific solution  in a specific mass point.
An example of such a solution area for one of its eight possible solutions of a single top pair event is plotted in Figure \ref{fig:singleevent_sol} (left).
The area provides  a boundary in the lower mass region  for the possible masses of top quark and W boson, as below these masses the event is not solvable.

%%%%%%%%%%%%%%%%%% F I G U R E %%%%%%%%%%%%%%%%%%%%%%%%%%%%%%%%%%%%%%%%%%%%%%
\begin{figure}[t!]
\includegraphics[scale=0.27]{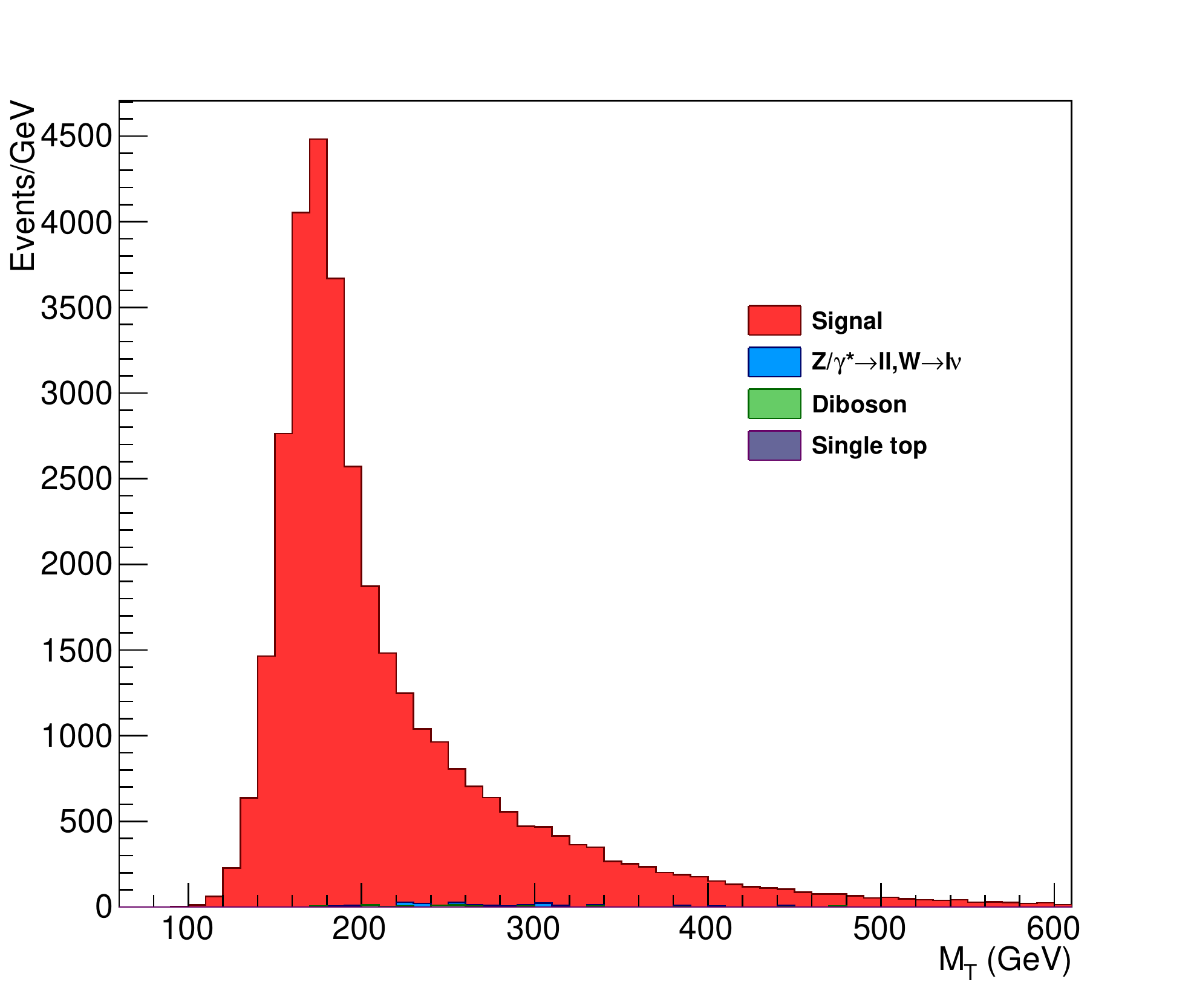}
\includegraphics[scale=0.27]{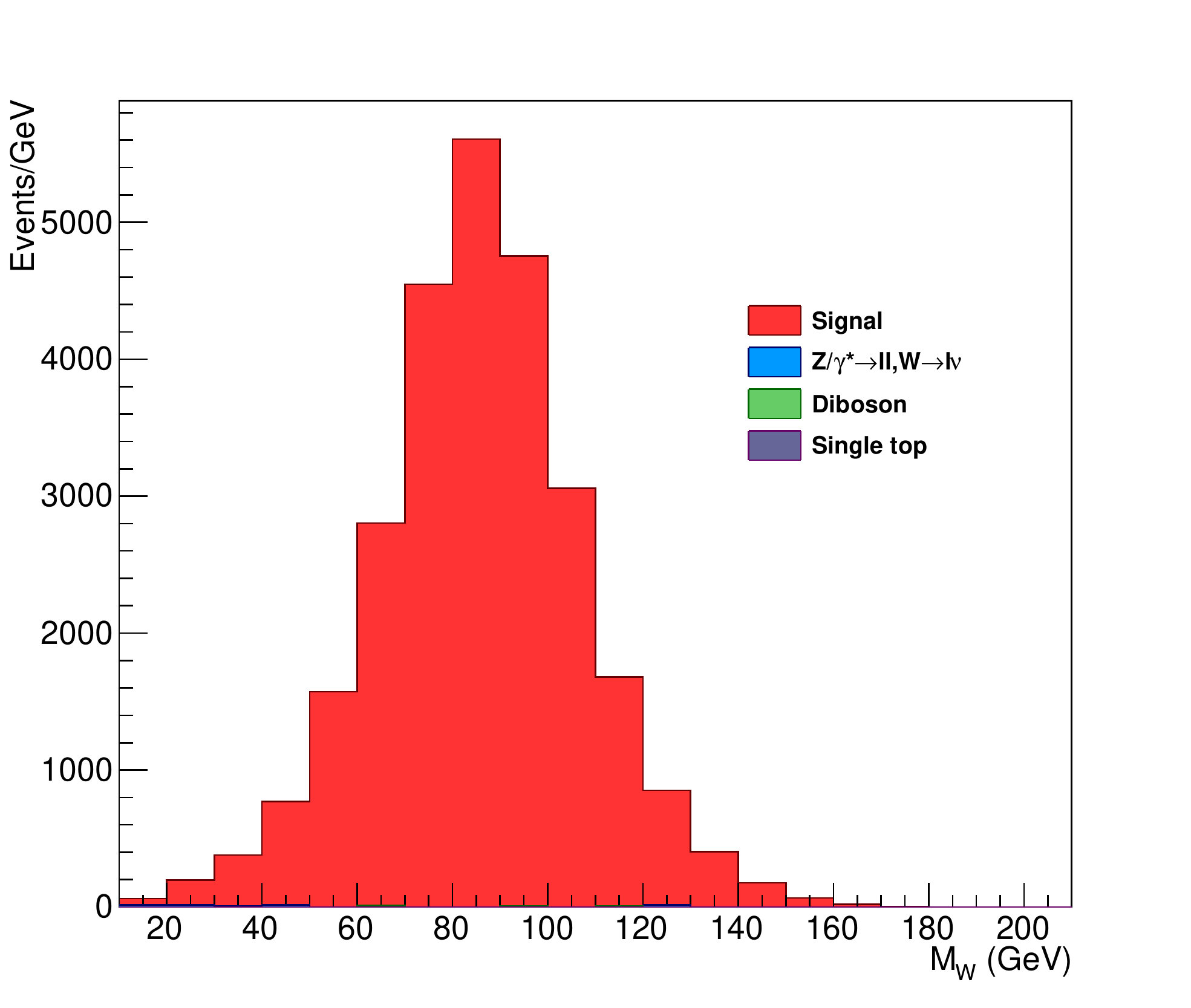}
\caption{
Reconstructed  $\mathrm{M_{T}}$ (left)  as well as $\mathrm{M_{W}}$ (right), for a sample of signal (top pairs) and background events corresponding to an intergrated luminosity of  50 $\mathrm{fb^{-1}}$ at  13 TeV.
}
\label{fig:toppairsmasses}
\end{figure}
%%%%%%%%%%%%%%%%%% F I G U R E %%%%%%%%%%%%%%%%%%%%%%%%%%%%%%%%%%%%%%%%%%%%%%

Due to the finite collission energy there is also an upper limit to the allowed masses produced.
The center of mass energy of the partons partipipating in the hard scattering has to be smaller than the LHC collision energy.
For a p-p collider this energy limit can be expressed fully by the parton distribution functions (PDFs) of the proton in the following way:
each solution allows full reconstruction of event kinematics, including the estimation of the energy E and $\mathrm{p_{Z}}$  momenta component of the $\mathrm{t\bar{t}}$ system.
These variables can be easily transformed to the fraction of beam energy of the two partons participating in the hard scattering ($\mathrm{ x_{1,2}=E\pm p_{Z}/(2 \sqrt{s} )) }$, $\mathrm{\sqrt{s}}$ being the center of mass energy).
So each parton with fraction  $\mathrm{x_{i}}$  can be assigned with a probability $\mathrm {F(x_{i}}$) to originate from a proton-proton collision.
By multiplying the probabilities of the two incoming partons, a weight per mass point can be assigned for each solution.
As there are more than one possible leading order parton-parton interactions
($\mathrm{u\bar{u}}$,  $\mathrm{\bar{u}u}$, $\mathrm{d\bar{d}}$, $\mathrm{\bar{d}d}$,  $\mathrm{gg}$), the
weights from all  possible combinations are summed to estimate a final event weight per solution and  mass point.
The weight can be written as $\mathrm{ \sum_{a,b}  F^{a}(x_{1},Q) F^{b}(x_{2},Q)}$, where the summation is over all possible parton combinations, $\mathrm {F^{a/b}(x,Q)}$ refers to the corresponding parton CT10 PDF set ~\cite{ct10} and $\mathrm{Q}$ is the momentum transfer (of the order of $\mathrm{M_{T}}$).
% with momentum transfer  $\mathrm{Q^{2}=m^{2}_{t}}$
For the estimation of the PDF values the LHAPDF-6.1.2 interface was used ~\cite{lhapdf}.

The PDF weight normalized to unit volume provides an upper bound for  the mass values of both $\mathrm{M_{T}}$ and $\mathrm{M_{W}}$.
The  solution area shown in Figure \ref{fig:singleevent_sol} (left) is weighted by the PDFs and the result is  plotted in  Figure \ref{fig:singleevent_sol} (right).
Each of the possible event solutions can produce such a distribution, so the maximum point of all distributions is the most likely to originate from a  proton-proton collision and is therefore taken as the $\mathrm{M_{T}}$ and $\mathrm{M_{W}}$ for this event.
It is interesting to mention that the prefered mass point  is not the one with the lowest  $\mathrm{M_{T}}$ and $\mathrm{M_{W}}$
values as one might have guessed from the fact that PDFs favour lower mass values.
Use of the solvability together with a matrix element weight which depends on the model has been proposed  for top quark mass estimation in a single mass dimension ~\cite{kondo1}-\cite{matrixweighting}.
This proposal has evolved to the matrix element weighting top mass measurement method in Tevatron ~\cite{D0topmass}, which has also been used at the LHC ~\cite{LHCtopmass}.
The proposal in this paper is to use the PDF weight to search for final states with two invisible particles in the 2-Dimensional mass plane of the unknown particles.
No  matrix elements are used so that the search is as model independent as possible.

Detector effects can change the momenta of the leptons and jets making a solvable event not-solvable.
In many cases solvability can be recovered by smearing the leptons and jets according to detector resolution.
For each initial event, N smeared events can be created by smearing the leptons and jets of the recorded event according to the detector resolutions.
For each of these smeared events, the same procedure as described above is followed: the solution area is weighted by  the PDFs.
The result for each solution is averaged over all N smeared events to form the final observable by the formula $\mathrm{ \sum_{i=1}^{N} \sum_{a,b}  F^{a}(x^{i}_{1},Q) F^{b}(x^{i}_{2},Q)}$, normalized to unit volume.
An example is given in Figure \ref{fig:singleevent_3D_300} for the same solution of the initial top pair event.
Again, among all solutions, the one with the maximum PDF weight is chosen as the final $\mathrm{M_{T}}$ and $\mathrm{M_{W}}$ estimation for this event. 
The above  procedure gives a single entry per event for each of the unknown masses.
Is is worth emphasizing that not all combinations/solutions are as likely to originate from a proton-proton collision and the parton distribution functions can distinguish one of them.
This might be  applicable to other cases with combinatoric backgrounds such as reconstruction of chains with visible particles.

\section{2D mass reconstruction - example topologies}

Several examples topologies for the method are described in this section starting from the benchmark top pairs in  the dilepton channel.
Next step is a search for anything decaying like dilepton top pairs, a generic topology concerning heavy top partners.
The 2-Dimensional mass reconstruction can be applied to other interesting searches with a single unknown mass as well as for a top mass measurement.
The identification of dilepton top pairs is another possible application as they constitute the most significant SM background in final states with missing energy.

\subsection{top pairs}

The method can be tested using existing SM particles such as top quark and W boson in the top pairs dilepton channel (Figure \ref{fig:feynman}, left). 
Simulated samples  corresponding to an  intergrated luminosity of 50 $\mathrm{fb^{-1}}$ for top-pairs and the background processes have been created.  
The latter consist of single boson (Drell-Yan, W+jets),  dibosons ($\mathrm{WW}$, $\mathrm{WZ}$, $\mathrm{ZZ}$) as well as single top samples.
By applying the method described in the previous section  in both signal (top-pairs)  and background events, the  2-Dimensional mass distribution  presented in Figure \ref{fig:2D_toppairs}  is created.
The top quark mass is shown in Figure \ref{fig:toppairsmasses}, using a range of 60-100 GeV for the W mass.
In a similar way, the W boson mass is presented in Figure \ref{fig:toppairsmasses} (right) for a range of 150-200 GeV of the top quark mass. 
It is worth mentioning that the W boson distribution has a resonance shape for a leptonic W decay.

%%%%%%%%%%%%%%%%%% F I G U R E %%%%%%%%%%%%%%%%%%%%%%%%%%%%%%%%%%%%%%%%%%%%%%
\begin{figure}[bh!]
\includegraphics[scale=0.091]{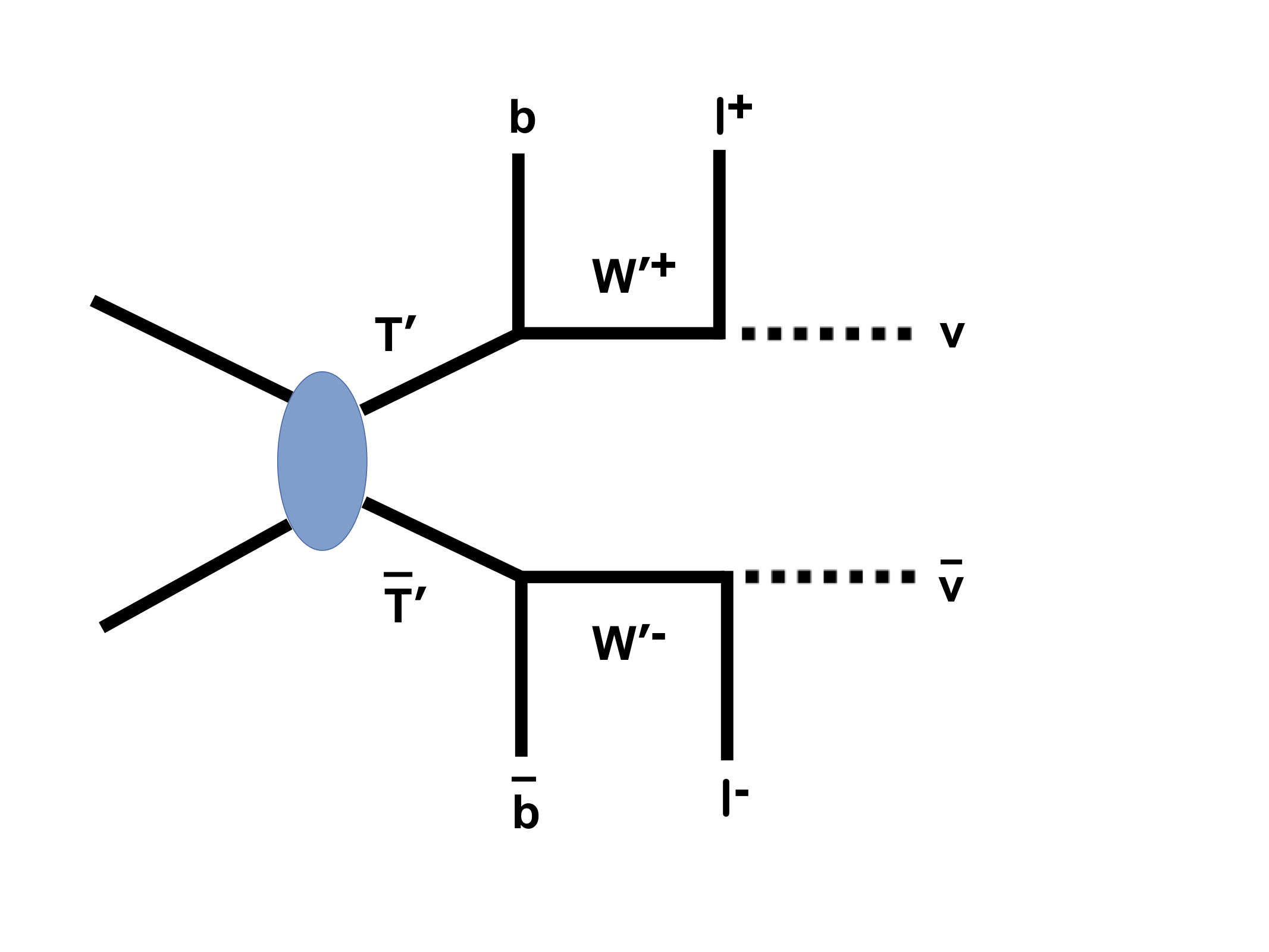}
\caption{
A topology for anything decaying like dilepton top pairs with both a new heavy top partner $\mathrm{T'}$ and a new heavy gauge boson $\mathrm{W'}$.
}
\label{fig:feynmananythingliketoppairs}
\end{figure}
%%%%%%%%%%%%%%%%%% F I G U R E %%%%%%%%%%%%%%%%%%%%%%%%%%%%%%%%%%%%%%%%%%%%%%

So without any a priori knowledge of their masses or of the underlying theory, both top quark and W boson can be observed simultaneously by assuming only the event topology. 
This is a proof of principle for the method, which can then be applied to searches for new hypothetical particles with  unknown masses.

%%%%%%%%%%%%%%%%%% F I G U R E %%%%%%%%%%%%%%%%%%%%%%%%%%%%%%%%%%%%%%%%%%%%%%
\begin{figure}[pt!]
\centering
\includegraphics[scale=0.35]{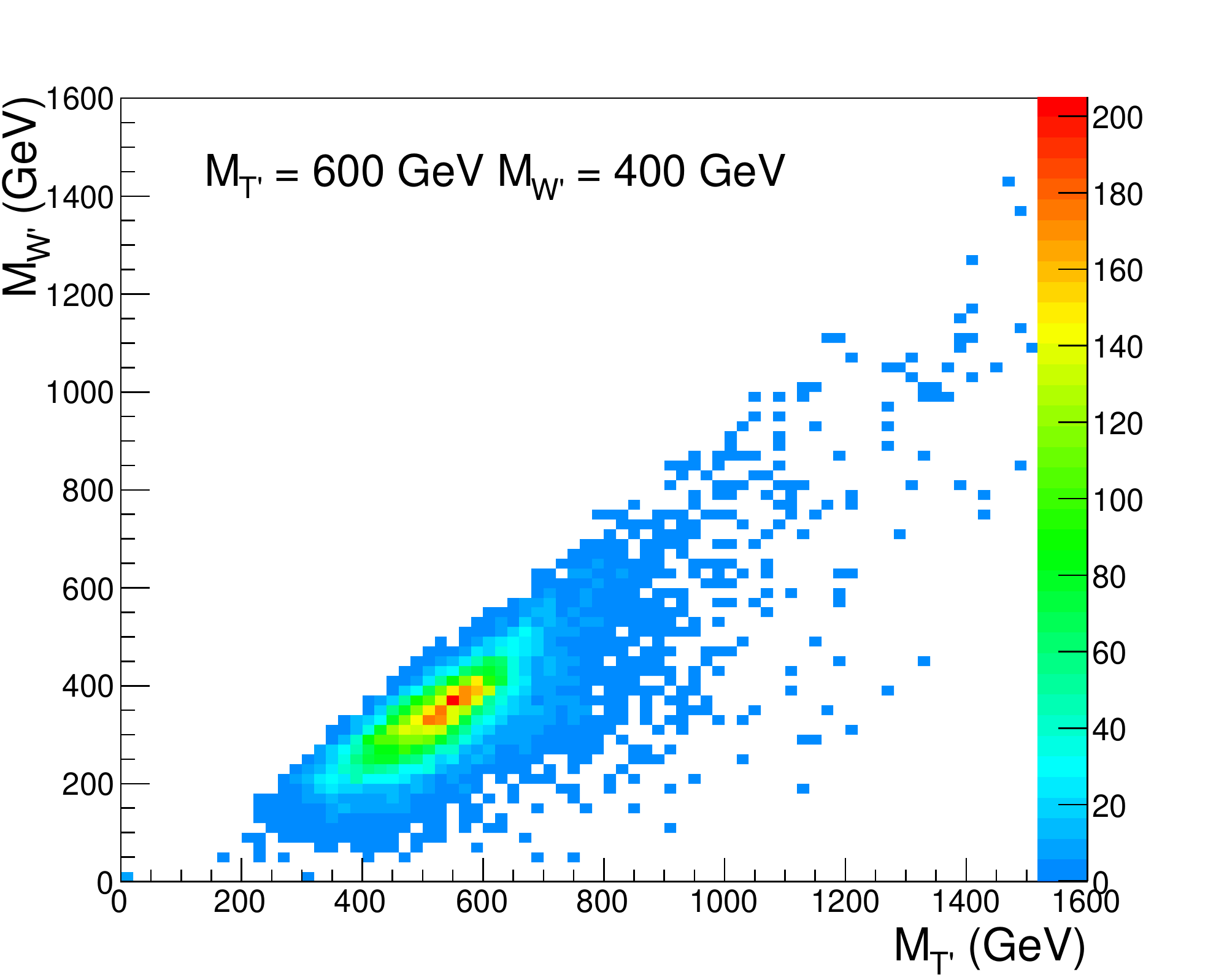}
\includegraphics[scale=0.35]{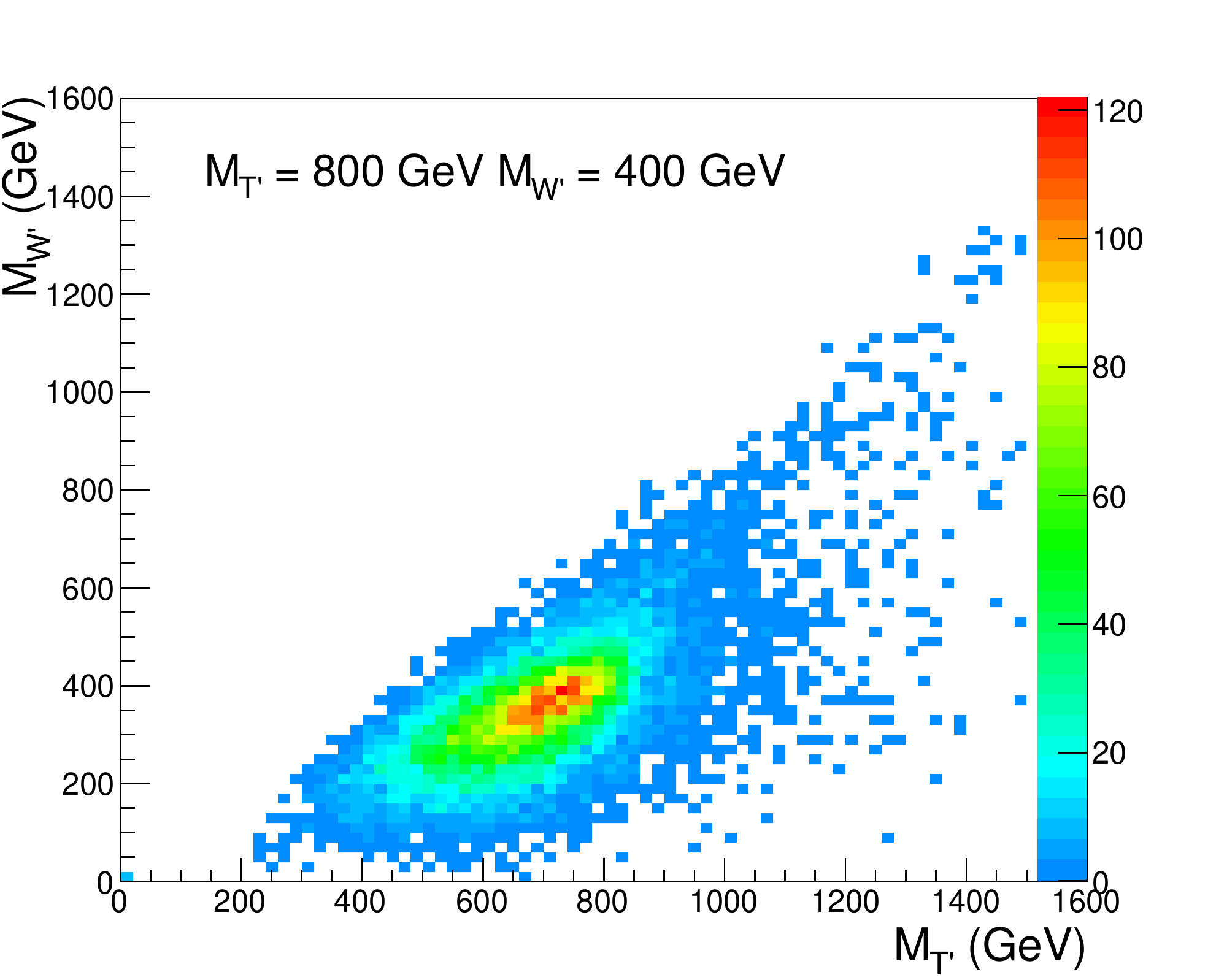}
\caption{
Reconstructed  signal distributions (littlest Higgs model) for $\mathrm{M_{T'} = 600}$ GeV, $\mathrm{M_{W'} = 400}$ GeV (left), as well as $\mathrm{M_{T'} = 800}$ GeV, $\mathrm{M_{W'} = 400}$ GeV (right).}
\label{fig:2DSignals1}
\vspace{2em}
\end{figure}
%%%%%%%%%%%%%%%%%% F I G U R E %%%%%%%%%%%%%%%%%%%%%%%%%%%%%%%%%%%%%%%%%%%%%%

%%%%%%%%%%%%%%%%%% F I G U R E %%%%%%%%%%%%%%%%%%%%%%%%%%%%%%%%%%%%%%%%%%%%%%
\begin{figure}[ph!]
\centering
\includegraphics[scale=0.35]{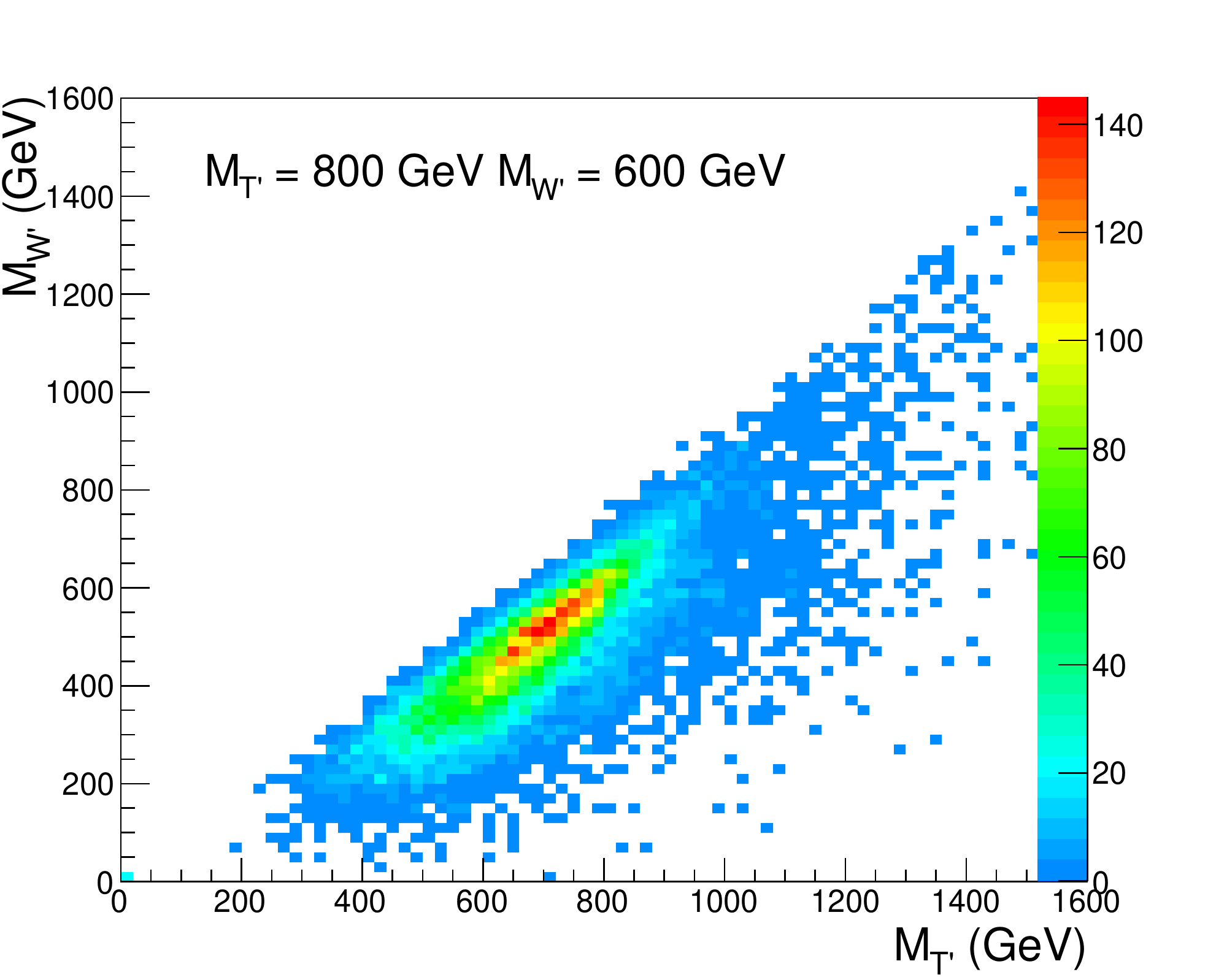}
\includegraphics[scale=0.35]{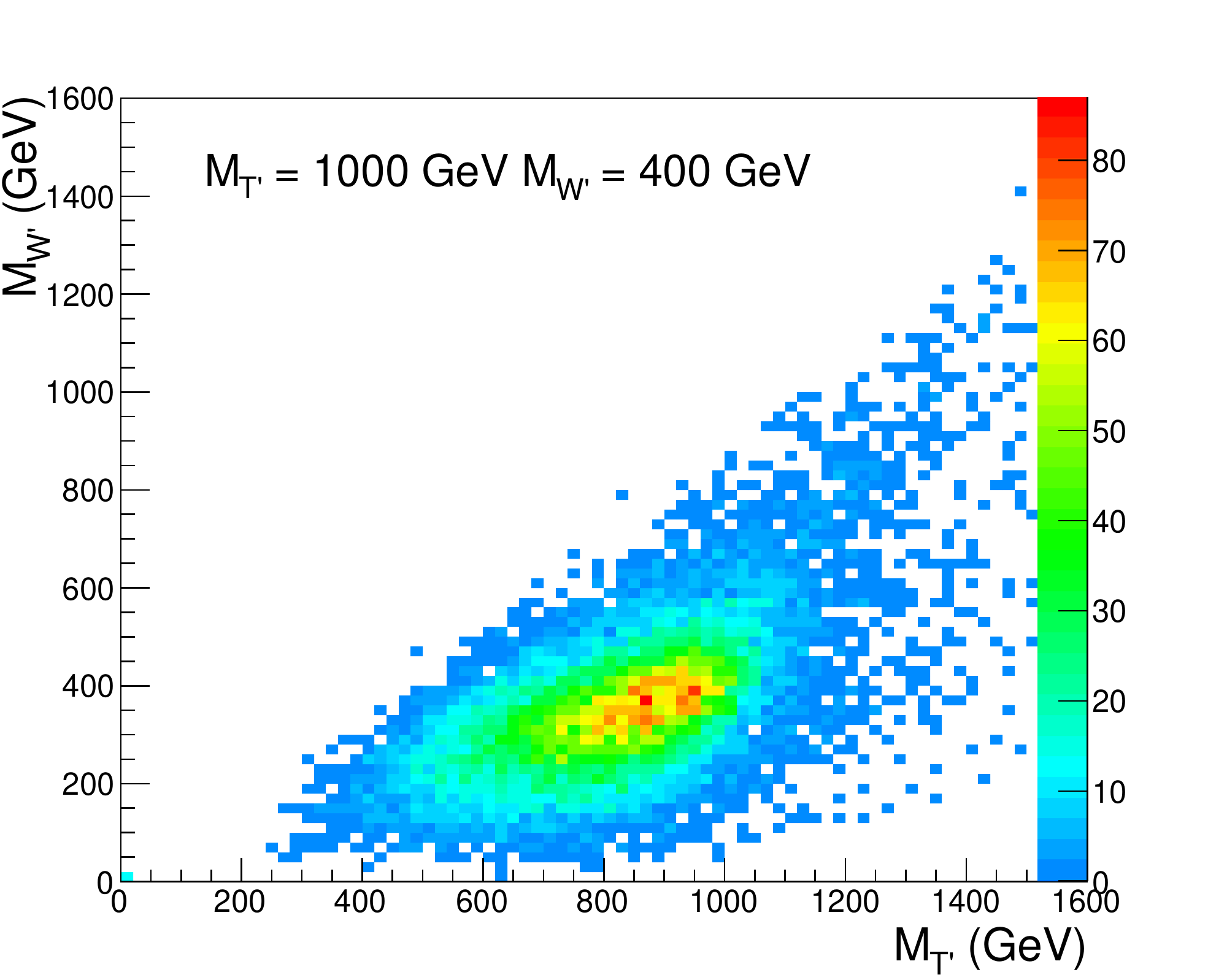}
\caption{Reconstructed  signal distributions (littlest Higgs model) for $\mathrm{M_{T'} = 800}$ GeV, $\mathrm{M_{W'} = 600}$ GeV (left), as well as $\mathrm{M_{T'} = 1000}$ GeV, $\mathrm{M_{W'} = 400}$ GeV (right).}
\label{fig:2DSignals2}
\vspace{2em}
\end{figure}
%%%%%%%%%%%%%%%%%% F I G U R E %%%%%%%%%%%%%%%%%%%%%%%%%%%%%%%%%%%%%%%%%%%%%%

%%%%%%%%%%%%%%%%%% F I G U R E %%%%%%%%%%%%%%%%%%%%%%%%%%%%%%%%%%%%%%%%%%%%%%
\begin{figure}[t!]
\centering
\includegraphics[scale=0.42]{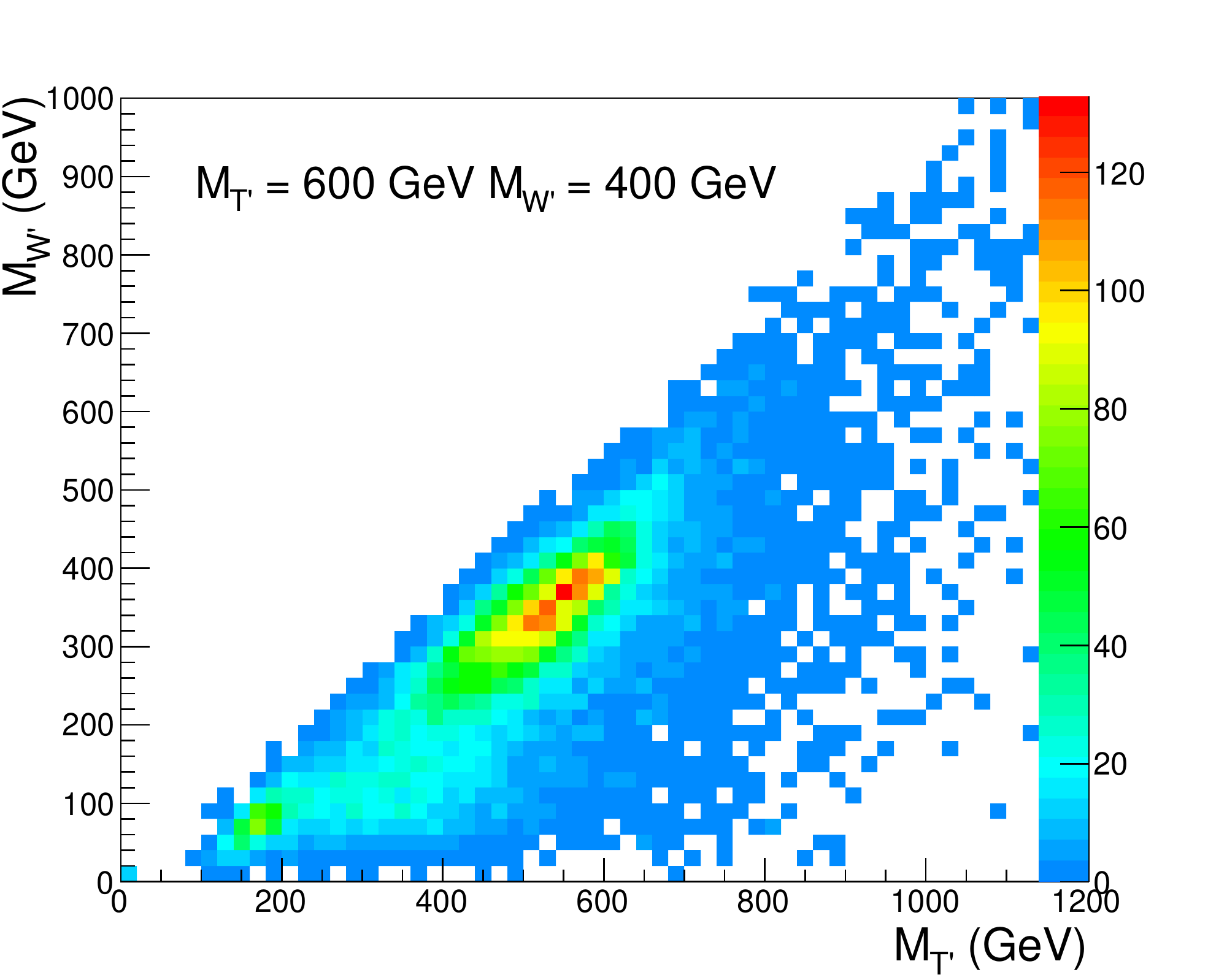}
\caption{Reconstructed  signal distributions for $\mathrm{M_{T'} = 600}$ GeV, $\mathrm{M_{W'} = 400}$ GeV (littlest Higgs model)  together with the expected background processes corresponding to an intergrated luminosity of  50 $\mathrm{fb^{-1}}$ at  13 TeV.}
\label{fig:2Dsignalplusbackground}
\vspace{2em}
\end{figure}
%%%%%%%%%%%%%%%%%% F I G U R E %%%%%%%%%%%%%%%%%%%%%%%%%%%%%%%%%%%%%%%%%%%%%%

\subsection{Search for anything decaying like dilepton top pairs}

The next step is the most generic search for anything decaying like dilepton top pairs ($\mathrm{pp \rightarrow T'\bar{T'} \rightarrow W'b W'b }$), in their dilepton final state in which both $\mathrm{W'}$ bosons decay leptonically ($\mathrm{W' \rightarrow e \slash \mu,\nu }$).
More specifically, this topology has two unknown particles, a new heavy top partner $\mathrm{T'}$ and a new heavy charged gauge boson $\mathrm{W'}$ (Figure \ref{fig:feynmananythingliketoppairs}).
It is a search for a heavy top partner in a quite generic topology.
The signal samples as already mentioned  are based on the littlest Higgs model.
The selection is the same as described in section 2 for the leptons and jets, with the missing energy requirement raised to 200 GeV.

The performance of the 2-Dimensional mass reco-nstruction can be seen in Figures \ref{fig:2DSignals1}, \ref{fig:2DSignals2}  for several signal samples.
Reconstructed signal together with the background procceses are presented in Figure \ref{fig:2Dsignalplusbackground}.
The lower region of the 2-Dimensional mass plane is populated by top pairs. 
Signal and background events live in different regions of the mass plane, resulting in a good discrimination between them.
The invariant mass of the $\mathrm{T'\bar{T'}}$ system is an interesting observable to monitor for possible new heavy neutral gauge bosons (e.g $\mathrm{pp \rightarrow Z' \rightarrow T'\bar{T'}}$).

\subsection{Top pair identification}

A possible application of the 2-Dimensional mass reconstruction is the identification of dilepton top pair events.
This is the most significant SM background for searches performed in final states with missing energy,
as it populates the tail of the missing energy related  observables used for discovery.
This is the same region where  possible signal events might exist.
The 2-Dimensional mass plane allows the significant suppression of dilepton top pairs by imposing constraints on the masses $\mathrm{M_{T}}$ and $\mathrm{M_{W}}$.

\subsection{Top mass measurement}

The 2-Dimensional mass reconstruction can also be used for a top mass measurement in the top pairs dilepton channel.
This is the cleanest channel as far as the background is concerned but it is considered to be more difficult in terms of the estimation of the dominant jet energy scale systematic effect.
This is due to the fact that both hadronic and semileptonic channels allow the simultaneous reconctruction of the W boson mass, which can further be used for the estimation of the jet energy scale uncertainty.
The 2-Dimensional mass reconstruction offers the opportunity to have  both masses $\mathrm{M_{T}}$, $\mathrm{M_{W}}$ in the relatively clean in terms of background dilepton channel, for a competitive top mass measurement.

%%%%%%%%%%%%%%%%%% F I G U R E %%%%%%%%%%%%%%%%%%%%%%%%%%%%%%%%%%%%%%%%%%%%%%
\begin{figure}[thb!]
\centering
\includegraphics[scale=0.28]{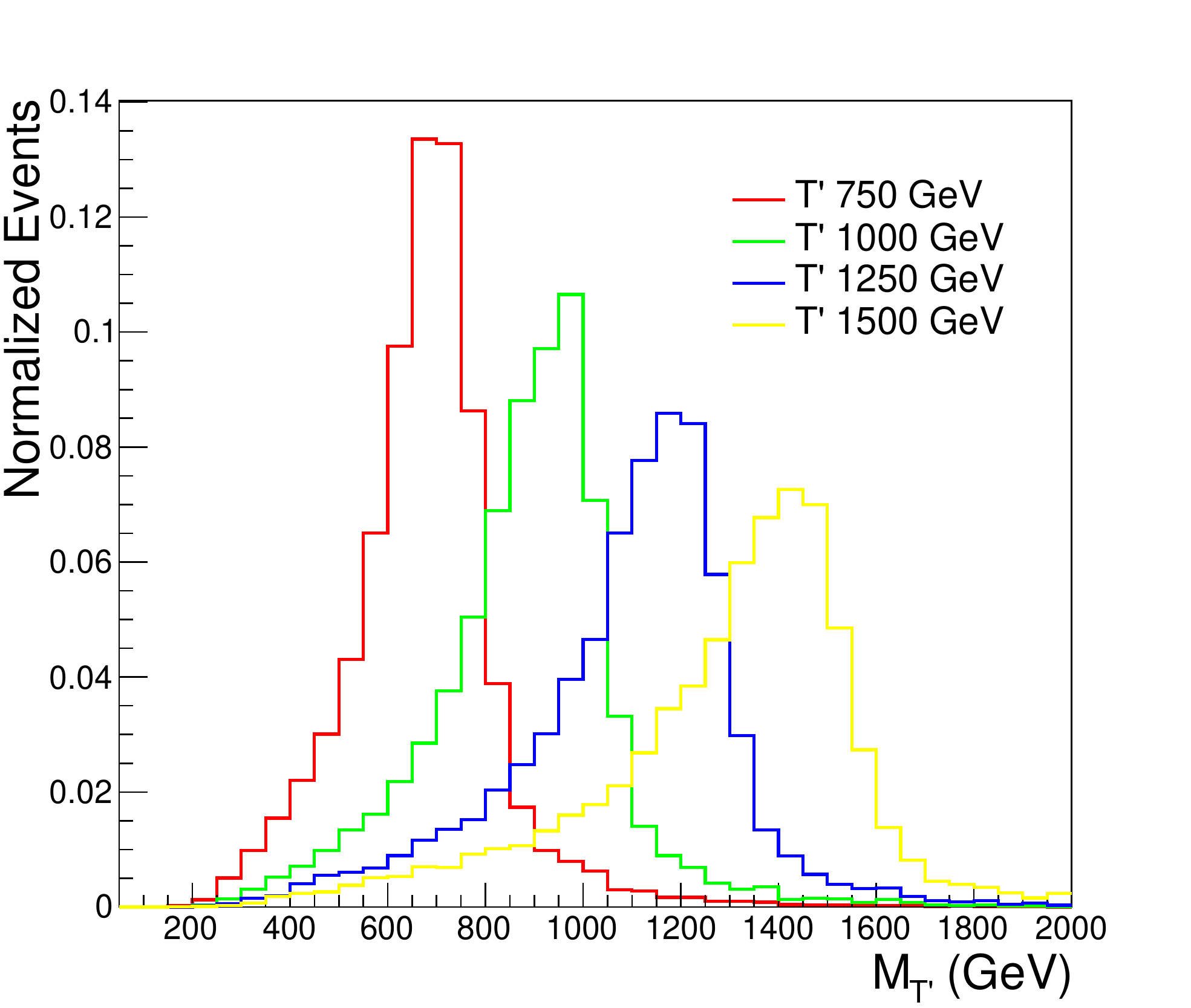}
\includegraphics[scale=0.28]{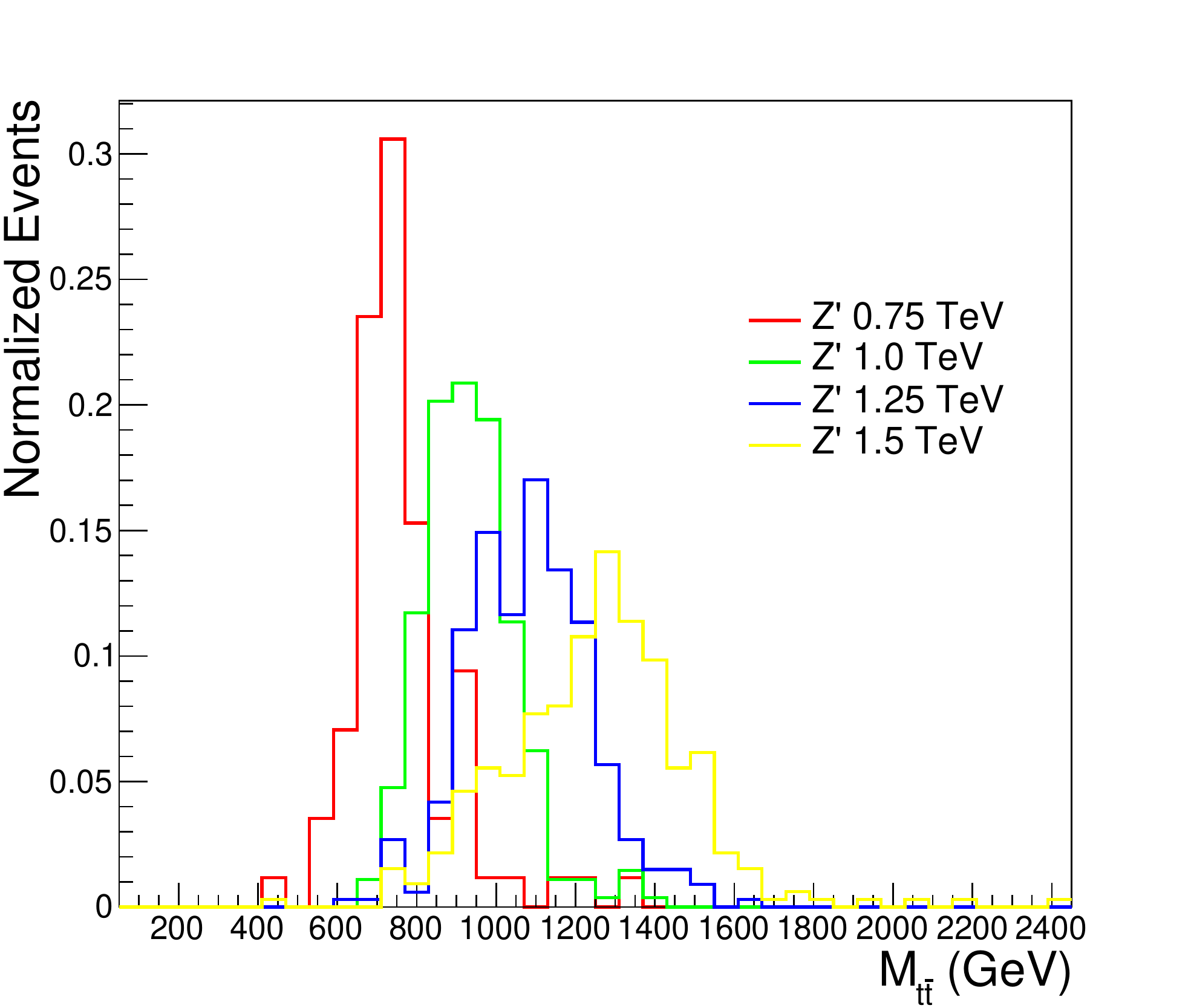}
\caption{The reconstructed masses for several $\mathrm{T'}$ (left) and $\mathrm{Z'}$ (right) signal samples.}
\label{fig:zprimetprime}
\vspace{2em}
\end{figure}
%%%%%%%%%%%%%%%%%% F I G U R E %%%%%%%%%%%%%%%%%%%%%%%%%%%%%%%%%%%%%%%%%%%%%

\subsection{Other topologies}

Other topologies with a single unknown mass can be reconstructed using the same method.
The classic search for a heavy top partner decaying as the SM top quark is such an example ($\mathrm{pp \rightarrow T'\bar{T'} \rightarrow W b W b})$. 
The results of the  $\mathrm{T'}$ reconstruction for several mass values can be seen in Figure  \ref{fig:zprimetprime} (left).
%Another possibility is to search for X particle instead of Y, where  X could be a charged higgs boson.
%The invariant mass distribution is an interesting observable to look for a heavy new particle as already mentioned.
%The search for a new heavy neutral gauge boson decaying to SM top pairs can also be performed ($\mathrm{pp \rightarrow Z' \rightarrow t\bar{t}})$.
Another applicable topology is the search for a new heavy neutral gauge boson decaying to SM top pairs ($\mathrm{pp \rightarrow Z' \rightarrow t\bar{t}})$.
The reconstructed $\mathrm{M_{Z'}}$ for several signal samples are shown in Figure  \ref{fig:zprimetprime} (right).

The most generic topology shown in Figure \ref{fig:feynman} (right) has an extra unknown mass compared to the diagram in Figure \ref{fig:feynmananythingliketoppairs}, as the neutrino is replaced by a massive WIMP.
In this case, the algorithm can give the masses of two unknown particles say X and Y provided that the third mass is given (e.g $\mathrm{M_{N}}$). 
For topologies concerning dark matter candidates this is an important issue as the third paricle is usually the LSP.
For this case, one option is to set the mass $\mathrm{M_{N}}$ equal to zero and then perform the 2-Dimensional mass reconstruction. 
In terms of discovery, it is still a bump hunt as the result is a peak in the 2-Dimensional mass plane of $\mathrm{M_{X}}$ and $\mathrm{M_{Y}}$,  but displaced to lower values. 
For our understanding of the BSM physics it is also important as two of its parameters ($\mathrm{M_{X}}$ and $\mathrm{M_{Y}}$) can be determined with respect to the third one (e.g $\mathrm{M_{N}}$).

\section{Conclusions}

Higgs boson search was a bumb hunt in an expected more or less region of invariant mass spectrum.
The collider, the experiments and the analysis were designed based on accurate simulation predictions.
The search for BSM physics is much harder. Well motivated theories with heavy top partners or dark matter candidates predict final states with large missing energy due to invisible particles.
In this cases, instead of a bump hunt the search is usually performed in the tail of a missing energy related observable.
Not only the shape of signal and background processes are similar, but also the discovery cannot give hints about the nature of new physics.

Mass space is the natural  space to search for new particles.
Mass observables do not require optimization or training.
%Signal resonances are concentrated in a small region, whereas background events have no reason to do the same.
This paper proposes to search for final states with two invisible particles in the 2-Dimensional mass space of the unknown particles.
The reconstruction is based on a PDF weight without any matrix elements, to be as model independent as possible for a given topology. 
Thus, the search is a bump hunt in more than one dimension, making signal discrimination from background processes an easier task.
In addition, reconstruction of the unknown masses can give valuable insights to what the new physics might be. 
%The search is as model independent as possible with only assumption  the decay topology.

Initially, the proof of principle is presented using the existing SM dilepton top pairs.
A generic search for anything decaying like dilepton top pairs with both a new heavy top partner and a new heavy gauge boson is used to show the application of the method in a typical topology with two invisible particles.
Top pair identification is an interesting application for searches using missing energy like observables.
The 2-Dimensional mass reconstruction can also be applied to many other topologies with one or two uknown masses as well as for a top mass measurement in the dilepton top pairs channel.
The method has already been used in CMS Run1 with many interesting topologies awaiting the next LHC Runs.

%%%%%%%%%%%%%%%%%%%%%%%%%%%%%%%%%%%%%%%%%%%%%%%%
%% BACKMATTER
%%%%%%%%%%%%%%%%%%%%%%%%%%%%%%%%%%%%%%%%%%%%%%%%

\appendix
\section*{Appendix}

The equations for the top pair system in the dilepton channel are the following: \\

\begin{tabular}{l l c}
  & $\mathrm{m^{2}_{t}}=\mathrm{(E_{b}+E_{l^{+}}+E_{\nu})^{2}}-\sum_{i=1}^{3} \mathrm{(p_{b_{i}}+p_{l^{+}_{i}}+p_{\nu_{i}})^{2}}$ & \\
  & $\mathrm{m^{2}_{\bar{t}} }  =  \mathrm{(E_{\bar{b}}+E_{l^{-}}+E_{\bar{\nu}})^{2} }  - \sum_{i=1}^{3}    \mathrm{(p_{\bar{b}_{i}}+p_{l^{-}_{i}}+p_{\bar{\nu}_{i}})^{2}}$ &  \\
  & & 
\end{tabular}

\begin{tabular}{l l c}
  & $\mathrm{m^{2}_{W^{+}}}   =  \mathrm{(E_{l^{+}}+E_{\nu})^{2}} -  \sum_{i=1}^{3}     \mathrm{(p_{l^{+}_{i}}+p_{\nu_{i}})^{2}}$ & \\
  & $\mathrm{m^{2}_{W^{-}}}   =  \mathrm{(E_{l^{-}}+E_{\bar{\nu}})^{2}}  - \sum_{i=1}^{3}   \mathrm{(p_{l^{-}_{i}}+p_{\bar{\nu}_{i}})^{2}}$ &
\end{tabular}

\begin{tabular}{c  c  c}
  & &  \\
  &  $  \mathrm{M_{ET_{x}}}     =   \mathrm{p_{\nu_{x}}+p_{\bar{\nu}_{x}}} $ &   $  \mathrm{E^{2}_{\nu} }   =   \mathrm{p^{2}_{\nu_{x}}}+  \mathrm{p^{2}_{\nu_{y}}} +  \mathrm{p^{2}_{\nu_{z}} }$  \\
  &  $  \mathrm{M_{ET_{y}}}   =  \mathrm{p_{\nu_{y}}+p_{\bar{\nu}_{y}}} $    &   $  \mathrm{E^{2}_{\bar{\nu}} }  =  \mathrm{p^{2}_{\bar{\nu}_{x}}}+  \mathrm{p^{2}_{\bar{\nu}_{y}}} + \mathrm{ p^{2}_{\bar{\nu}_{z}}}$

\end{tabular}



\bibliographystyle{aipproc}   % if natbib is available

%\bibliography{sample}

\IfFileExists{\jobname.bbl}{}
 {\typeout{}
  \typeout{******************************************}
  \typeout{** Please run "bibtex \jobname" to optain}
  \typeout{** the bibliography and then re-run LaTeX}
  \typeout{** twice to fix the references!}
  \typeout{******************************************}
  \typeout{}
 }

%\end{document}

%%%%%%%%%%%%%%%%%%%%%%%%%%%%%%%%%%%%%%%%%%%
%% The following lines show an example how to produce a bibliography
%% without the help of the BibTeX program. This could be used instead
%% of the above.
%%%%%%%%%%%%%%%%%%%%%%%%%%%%%%%%%%%%%%%%%%%

\end{document}